\documentclass[table]{ccjnl}
\usepackage{lipsum,amsmath}
\usepackage{cuted}
\usepackage{color}
\usepackage{multirow} 
\graphicspath{{figures/}}

\title{Abnormal Signal Recognition with Time-Frequency Spectrogram: A Deep Learning Approach}
\author{Tingyan Kuang\inst{1}, Huichao Chen\inst{1}, Lu Han\inst{1}, Rong He\inst{1}, [Wei Wang\inst{1}, Guoru Ding\inst{2}\corinfo{Kungtingyan@nuaa.edu.cn}
}
\receiveddate{xxx}
\reviseddate{xxx}
\Editor{xxx}

\address[1]{College of Electronic and Information Engineering, Nanjing University of
Aeronautics and Astronautics, Nanjing, 211106, China, Key Laboratory of Dynamic Cognitive System of Electromagnetic Spectrum Space, Ministry of Industry and Information Technology}
\address[2]{College of Communications Engineering, Army Engineering University, Nanjing, 211106, China}

\begin{document}

\maketitle

\begin{abstract}
With the increasingly complex and changeable electromagnetic environment, wireless communication systems are facing jamming and abnormal signal injection, which significantly affects the normal operation of a communication system. In particular, the abnormal signals may emulate the normal signals, which makes it very challenging for abnormal signal recognition. In this paper, we propose a new abnormal signal recognition scheme, which combines time-frequency analysis with deep learning to effectively identify synthetic abnormal communication signals. Firstly, we emulate synthetic abnormal communication signals including seven jamming patterns. Then, we model an abnormal communication signals recognition system based on the communication protocol between the transmitter and the receiver. To improve the performance, we convert the original signal into the time-frequency spectrogram to develop an image classification algorithm. Simulation results demonstrate that the proposed method can effectively recognize the abnormal signals under various parameter configurations, even under low signal-to-noise ratio (SNR) and low jamming-to-signal ratio (JSR) conditions.
\end{abstract}
\section{Introduction}
\label{Introduction}
The electromagnetic spectrum has become a national strategic resource. Therefore, the effective use and management of the radio spectrum is significant \cite{5061975,9193946,7147778,2014Spatial,8741198}. The management of radio spectrum resources should not only monitor the occupancy of spectrum resources, but also supervise the working status of authorized signals. The open nature of radio waves makes them susceptible to interference and illegal use, causing normal communication systems to be disrupted or even interrupted, thereby threatening social and national security. In wireless communications, jamming represents the most severe security threat in the field of tactical radio systems. Jammers interfere with the communication links to interrupt the legally transmitted signal between the primary user (PU) and secondary users (SUs) \cite{7416819}. Therefore, we can discover illegal signals through abnormal spectrum data, which is beneficial to our information protection, especially in key areas such as airports and borders.

The signal anomaly detection in wireless communication is very different from other anomaly detection tasks, mainly reflected in two aspects: 1) The diversity of anomaly signal types makes it impossible to mark a large amount of anomaly signal data. 2) Early radio signal identification technology uses various spectrum acquisition equipment to collect radio signals. Then, experts use time-domain waveform diagrams or spectrograms to analyze signal characteristics. In this way, the results of monitoring are subjective, and manual monitoring is time-consuming. In addition, the complexity and quantity of electromagnetic environment data increase the difficulty of manual feature extraction. Traditional machine learning algorithms, such as Naive Bayes, neural networks, decision trees, etc., have been applied in the field of radio modulation recognition \cite{7920749}. However, only machine learning algorithms were used to classify signal features manually extracted. Therefore, we exploit deep learning algorithms to automatically extract the characteristics of communication signals, which can avoid experience based manual feature extraction and improve the ability to identify communication signals in a complex electromagnetic environment.

Abnormal detection has been investigated in the literature. Liu et al. proposed the dynamic spectrum access (DSA) anomaly detection method, which utilizes a distributed power measurement method of collaborative sensing for anomaly detection \cite{5061975}. Similarly, the Hidden Markov Model (HMM) spectral amplitude probability was used to detect jamming on the channel of interest again in the DSA domain in \cite{7428581}. Besides, neural networks have been used to detect anomalies in a single category or multiple categories. Hodge and Austin conducted an in-depth study on the anomaly detection technology developed in the field of machine learning and statistics \cite{2004A}. Agyemang et al. extensively discussed anomaly detection techniques for digital and symbolic data \cite{2006A}. Markou and Singh introduced the novelty detection techniques of neural networks and statistical methods respectively \cite{2003Novelty}.

Nowadays, with the vigorous development of artificial intelligence in all walks of life, researchers have done a lot of work in trying to use machine learning models to improve the efficiency of abnormal signal recognition. In \cite{6037247}, Lin et al. proposed a radar signal recognition neural network classification algorithm with a recognition rate that conforms to the actual situation. A recurrent neural network (RNN) anomaly detector based on predictive modeling of raw IQ data has been proposed in \cite{2016arXiv161100301O}. However, it is still not sufficiently automated and versatile for actual anomaly detection. Tandiya et al. proposed a wireless system anomaly detection method based on time-frequency spectrograms to monitor and analyze radio frequency spectrum activities \cite{8403654}. The authors in \cite{7949002} proposed a reliable method based on time-frequency analysis and deep learning to recognize the cognitive radio waveform. In \cite{4200706}, Lunden et al. comprehensively extracted various signal statistical features, combined with Wigner-Ville distribution (WVD) transformation to extract features, and identified LFM signals, P1, P2, P3, P4 codes, and Frank signals. The simulation results show that this method is automatic and efficient.

In this paper, we combine time-frequency analysis with deep learning to effectively recognize abnormal communication signals. The contributions can be summarized as follows:
\begin{itemize}
\item We first emulate synthetic abnormal communication signals including seven jamming patterns. Then, we build an abnormal communication signals recognition model based on the communication protocol between the transmitter and the receiver.
\item We develop an image classification algorithm based on a convolutional neural network (CNN). Specifically, we convert the original signal into a time-frequency spectrogram for a CNN. Then, the communication signal recognition problem is transformed into an image classification problem. In this paper, the convolutional neural networks VGG16, ResNet50, and InceptionV3 are adjusted and compared.
\item We conduct simulations to evaluate the performance of the proposed approach. It is shown that compared to the KNN and Naive Bayes algorithms, the proposed algorithm based on deep learning of time-frequency spectrograms can yield higher accuracy in the recognition of abnormal signals. Besides, we prove the model proposed has the capability to detect abnormalities when a new type of abnormal signal appears.
\end{itemize}

The remainder of the paper is organized as follows. The system model is described in Section II. The time-frequency analysis method for abnormal signal recognition is introduced and compared in Section III. Section IV presents the deep learning-based abnormal signal recognition scheme with time-frequency spectrograms. Simulation results under various parameter configurations are described in Section V. Section VI gives the conclusion and future work. 

\section{System Model}
\label{System}

\subsection{Normal Signals and Abnormal Signals}
\begin{figure}[htb]
    \centering
    \includegraphics[width=0.48\textwidth]{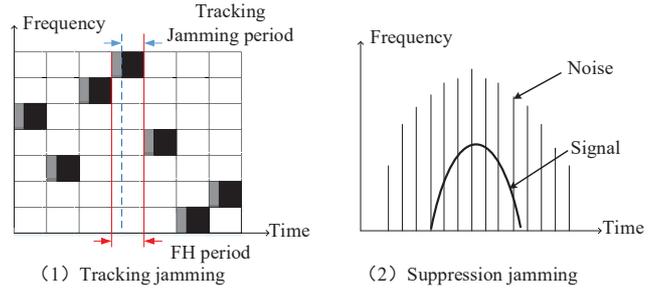}
    \caption{Tracking jamming and suppression jamming.}
	\label{fig2}
\end{figure}

There are two ways to define abnormal signals. One is for unauthorized signals, that is, signals sent from illegal transmitters without the authorization of radio regulatory agencies. The other is an authorized signal, but it is subject to external malicious jamming or failures that occurred on the receiver side during transmission. Generally, communication jamming is divided into two categories: suppression jamming and deception jamming. The typical communication jamming includes single-tone jamming, multi-tone jamming, linear frequency sweep jamming, and noise FM jamming. The abnormal signal identification usually firstly detects whether the abnormal signal exists, then identifies the type of communication jamming signal, and finally takes corresponding measures for the specific abnormal type. Among them, anomaly recognition technology is the prerequisite and key to anti-interference, and it also determines whether the corresponding anti-interference measures are effective. Therefore, we considered the recognition of the four communication jamming signals mentioned above. In this paper, for non-stationary signals, abnormal signals can be described as the following:
\begin{itemize}
\item The signal is legal, but the parameters such as the bandwidth and the center frequency of the signal are not within the specified range.
\item This is one synthetic anomaly, where legal signals coexist with jamming signals and the jamming signal parameters are legal (e.g. the signal with tracking jamming and suppression jamming in Figure~\ref{fig2}).
\end{itemize}

In this paper, we focus on the second type of anomaly. Then, the state of communication signals is classified into nine cases such as two normal cases and seven abnormal cases. We emulate two normal communication signals i.e., (i) \emph{FH}: the frequency-hopping signal (FH) generated by pseudo-random code can be expressed as 
\begin{equation}
S_0\left( t \right)=m\left( t \right)\text{cos}\left( {{w}_{j}}t+{{\theta }_{j}} \right), \label{eq1}
\end{equation}
where $m\left( t \right)$ denotes the information data signal; $w_{j}$ and $\theta_{j}$ denote the angular frequency and phase in the time interval $jT \leq t < \left( j + 1 \right)T$; $T$ denotes the hopping time interval. In this paper, for the convenience of simulation, we assume that the form of the signal is given by
\begin{equation}
S\left( n \right)=\left\{ \begin{matrix}
   \begin{matrix}
   \cos \left( 2\pi {{f}_{1}}{{T}_{s}}n+{{\theta }_{1}} \right),~~~~~~~~~~~~~~0\le n\le N,  \\
   \cos \left( 2\pi {{f}_{2}}{{T}_{s}}n+{{\theta }_{2}} \right),~~~~N+1\le n\le 2N,  \\
\end{matrix}  \\
   \begin{matrix}
   \cos \left( 2\pi {{f}_{3}}{{T}_{s}}n+{{\theta }_{3}} \right),~~2N+1\le n\le 3N,  \\
   \cos \left( 2\pi {{f}_{4}}{{T}_{s}}n+{{\theta }_{4}} \right),~~3N+1\le n\le 4N,  \\
\end{matrix}  \\
   \begin{matrix}
   \cos \left( 2\pi {{f}_{5}}{{T}_{s}}n+{{\theta }_{5}} \right),~~4N+1\le n\le 5N,  \\
   \cos \left( 2\pi {{f}_{6}}{{T}_{s}}n+{{\theta }_{6}} \right),~~5N+1\le n\le 6N,  \\
\end{matrix}  \\
\end{matrix} \right. \label{eq11}
\end{equation}
where $T_s$ denotes the sampling period; $f_1$, $f_2$, $f_3$, $f_4$, $f_5$, $f_6$ denote the frequency point of the frequency hopping signal, respectively; $N$ denotes the number of sampling points in $T$.

(ii) \emph{BPSK}: the BPSK signal with the random carrier frequency can be expressed as
\begin{equation}
S_0\left( t \right) = \left( {\sum\limits_{n}{a_{n}g\left( t - nT_{s} \right)}} \right)\text{cos}w_{c}t, \label{eq2}
\end{equation}
where $g\left( t \right)$ denotes a rectangular pulse with a pulse width of $T_s$. The statistical properties of $a_n$ can be expressed by
\begin{equation}
a_{n} = \left\{ \begin{matrix}
{1,~~~~~~with~~Prob.P} \\
{- 1,~~with~~Prob.1 - P} \\
\end{matrix} \right., \label{eq2.1}
\end{equation}
where $P$ denotes the probability of sending the binary symbol '0', and $1-P$ denotes the probability of sending the binary symbol '1'. When the sent binary symbols are 0 and 1, respectively, $S_0\left( t \right)$ takes 0 and $\pi$ phase correspondingly.

Similarly, seven jamming pattern are emulated, which are given by (i) \emph{Tracking-jamming}: jamming signals form the tracking jamming at the same frequency as the FH signal, and the power of the jamming signal is much larger than the FH signal. It can be expressed by
\begin{equation}
\begin{aligned}
&~~~~~~S(n)=\\
&\left\{\begin{array}{ll}
\mathrm{A}_{1}  \cos \left(2 \pi f_{1} T_{s} n+\theta_{1}\right),& m \leq n \leq N, \\
\mathrm{A}_{2} \cos \left(2 \pi f_{2} T_{s} n+\theta_{2}\right),& m+N+1 \leq n \leq 2N, \\
\mathrm{A}_{3} \cos \left(2 \pi f_{3} T_{s} n+\theta_{3}\right),& m+2N+1 \leq n \leq 3N, \\
\mathrm{A}_{4} \cos \left(2 \pi f_{4} T_{s} n+\theta_{4}\right),& m+3N+1 \leq n \leq 4N, \\
\mathrm{A}_{5} \cos \left(2 \pi f_{5} T_{s} n+\theta_{5}\right),& m+4N+1 \leq n \leq 5N, \\
\mathrm{A}_{6} \cos \left(2 \pi f_{6} T_{s} n+\theta_{6}\right),& m+5N+1 \leq n \leq 6N,
\end{array}\right. \label{eq3}
\end{aligned}
\end{equation}
where $A$ denotes the amplitude; $m$ denotes the tracking delay.

(ii) \emph{Sweeping}: linear frequency sweeping jamming on the BPSK signal and the center frequency of the jamming signal change with the BPSK signal, 
\begin{equation}
S_j\left( t \right)=A\text{exp}\left( j\left( 2\pi {{f}_{0}}t+\pi k{{t}^{2}}+\varphi  \right) \right),0\le t\le T, \label{eq4}
\end{equation}
where $A$ denotes jamming amplitude; $f_{0}$ denotes the initial frequency of the jamming signal; $k$ denotes the frequency modulation coefficient; $\varphi$ denotes the initial phase; $T$ denotes duration.

(iii) \emph{Noise-FM}: the noise FM jamming with a certain bandwidth and high power is superimposed on the BPSK signal. The jamming is given by
\begin{equation}
S_j\left( t \right) = U_{j}\text{exp}\left( {j2\pi f_{j}t + 2\pi K_{FM}{\int_{0}^{t}{u\left( t \right)d\tau}}} \right), \label{eq5}
\end{equation}
where $U_{j}$ denotes the amplitude; $f_{j}$ denotes the center frequency; $K_{FM}$ denotes the frequency modulation coefficient; $u\left( t \right)$ denotes narrow-band Gaussian white noise, with a mean value of 0 and a variance of $\sigma_{n_j}^{2}$; $\int_{0}^{t}{u\left( \tau \right)d\tau}$ denotes Wiener process, subject to $N\left( 0, \sigma_{n}^{2}\tau \right)$ distribution.

(iv) \emph{Pulse}: the pulse jamming with random time transmission on the given band, which can be expressed as 
\begin{equation}
S_j\left( t \right) = A{\sum\limits_{k = 1}^{N}{\text{Rect}\left( t - kT_{r} \right)}}, \label{eq6}
\end{equation}
where $Rect\left( t \right)$ denotes the rectangular pulse with a pulse width of $T$; $T_{r}$ denotes the pulse period; $k$ denotes the number of pulses; $A$ denotes the pulse amplitude.

(v) \emph{Single-tone}: the single-tone jamming can emit high-power jamming to the BPSK signal and can be expressed by 
\begin{equation}
S_j\left( t \right)=\sqrt{2{{P}_{j}}}\text{cos}\left( 2\pi {{f}_{j}}t+\varphi  \right), \label{eq7}
\end{equation}
where $P_{j}$ denotes the power; $f_{j}$ denotes the frequency; $\varphi$ denotes the initial phase.

(vi) \emph{Multi-tone}:  the multi-tone jamming is composed of multiple tones. Usually, each frequency point of the multi-tone jamming is randomly distributed in a specific frequency band. It coexists with the BPSK signal and can be expressed as
\begin{equation}
S_j\left( t \right)=\underset{i=1}{\overset{N}{\mathop \sum }}\,\sqrt{2{{P}_{j}}/N}\text{cos}\left( 2\pi {{f}_{j,i}}t+\varphi  \right), \label{eq8}
\end{equation}
where $N$ denotes the number of tones; $f_{j,i}$ denotes the frequency of the $i$th tone of the multi-tone jamming; $\varphi$ denotes the initial phase of the $i$th tone. 

(vii) \emph{Comb-spectrum}:  the comb spectrum jamming on the BPSK signal is composed of multiple sub-jamming that undergo the same modulation. It is a set of narrow-band jamming modulated in a certain way at a series of frequency points. Each sub-jamming is superimposed in the time domain and separated in the frequency domain. The jamming spectrum presents a comb-like distribution. It can be expressed by
\begin{equation}
S_j\left( t \right)=\underset{k}{\mathop \sum }\,\sqrt{2{{P}_{k}}}\text{exp}\left[ j\left( 2\pi {{f}_{k}}t+\alpha \text{ }\!\!\Delta\!\!\text{ }fsin\left( \beta \pi t \right) \right) \right], \label{eq9}
\end{equation}
where $k$ denotes the number of comb teeth; $P_k$ denotes the power of each component; $f_k$ denotes the center frequency of each component; $2\Delta f$ denotes the bandwidth of each component; $\alpha$ and $\beta$ are the modulation parameters of each component. 

\subsection{Abnormal Signal Recognition Model}
\label{Abnormal}
\begin{figure}[htb]
    \centering
    \includegraphics[width=0.48\textwidth]{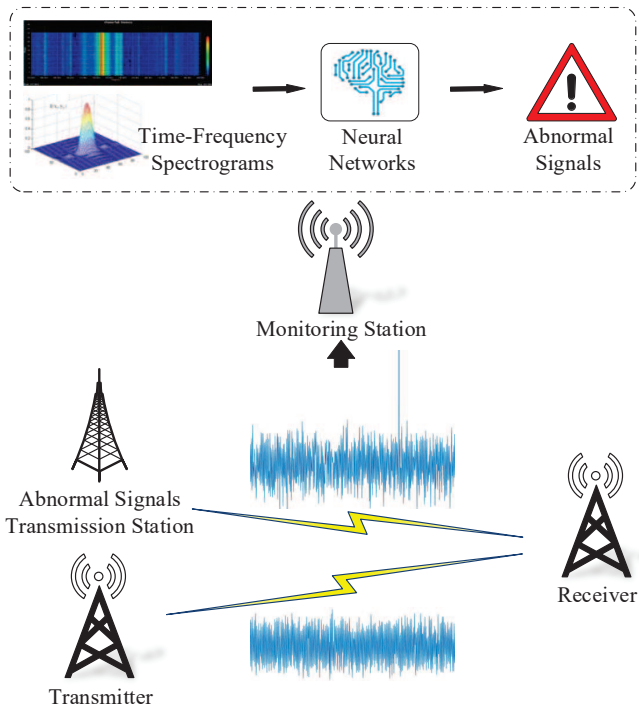}
    \caption{Abnormal communication signal recognition model.}
	\label{fig3}
\end{figure}

Since the communication signals are susceptible to environmental interference, it is complex and variable. Therefore, the key to designing abnormal signal recognition is to select a model trained with historical signal data. Then, we exploit this model to recognize abnormal signals that do not belong to normal signals. We consider abnormal communication signal recognition based on the general model of the communication system, and the overall system model is shown in Figure~\ref{fig3}.

The monitoring station performs detection during the entire communication process. Firstly, the monitoring station conducts time-frequency analysis with the received signal data, and then sends the time-frequency spectrogram obtained by time-frequency analysis into the convolutional neural network for identification and judgment. The convolutional neural network is a model that is pre-trained with historical data and has learned the features of signals.

In this paper, the abnormal signal is modeled as follows:
\begin{equation}
S_{1}\left( t \right) = S_{0}\left( t \right) + S_j\left( t \right), \label{eq12}
\end{equation}
where $S_{0}\left( t \right)$ denotes the normal signal mentioned above; $S_j\left( t \right)$ denotes the jamming mentioned above.

Concurrently, for such a synthetic abnormal signal, we also set the jamming-to-signal ratio (JSR), which can be expressed as
\begin{equation}
JSR = 10{log}_{10}\frac{P_{j}}{P_{s}}, \label{eq12.1}
\end{equation}
where $P_j$ denotes the power of the jamming; $P_{s}$ denotes the power of the normal signal.

Then, the received signal is modeled as follows:
\begin{equation}
r_{0}\left( t \right) = S_{0}\left( t \right) \times C_{0}\left( t \right) + n_{0}\left( t \right), \label{eq13}
\end{equation}
\begin{equation}
r_{1}\left( t \right) = S_{1}\left( t \right) \times C_{1}\left( t \right) + n_{1}\left( t \right), \label{eq14}
\end{equation}
where $r_{0}\left( t \right)$ denotes the normal signal received; $r_{1}\left( t \right)$ denotes the abnormal signal received; $C_{0}\left( t \right)$ and $C_{1}\left( t \right)$ denote the channel of the normal signal and the channel of the abnormal signal, respectively; $n_{0}\left( t \right)$ and $n_{1}\left( t \right)$ are modeled as i.i.d. Gaussian distribution with zero mean and variance $\sigma_{n}^{2}$. 

\section{Time-frequency Analysis for Abnormal Signal Recognition}
Generally speaking, there are two ways to represent signals: the waveform in the time domain and the spectrum in the frequency domain. The former is more intuitive, while the latter often uses Fourier transform to transform the signal from the time domain to frequency domain, to express its amplitude and phase. Although we can see the frequency composition of the signal through frequency domain analysis, it does not show the change in frequency. Only analyzing the global characteristics of the signal in the time domain or the frequency domain will ignore their local information and fail to meet the requirements of signal processing and analysis. We need to know not only the frequency of the signal but also how the frequency changes over time. Thus, one-dimensional solutions seem not to be sufficient, and one has to consider two-dimensional functions.

By solving the two-dimensional joint function of time and frequency to obtain the local information of the signal, the basic idea of time-frequency analysis comes from this. Time-frequency analysis is a non-stationary signal processing tool, which can be used to extract the characteristic information of a communication signal at a specific time and a specific frequency. Time-frequency analysis accurately reflects the relationship between the frequency of the signal over time, to help us find the features of non-stationary signals from the perspective of the time-frequency domain. 

\begin{figure}[htb]
    \centering
    \includegraphics[width=0.48\textwidth]{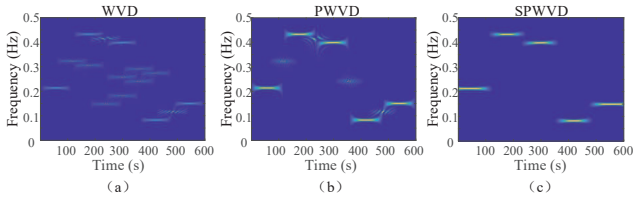}
    \caption{Illustration of three time-frequency spectrograms of FH signal.}
	\label{fig4}
\end{figure}

\subsection{Smooth Pseudo Wigner-Ville Distribution}
Smoothing can make the time-frequency features of multi-component signals more obvious. Therefore, in this paper, the time-frequency spectrogram is obtained by the smooth pseudo Wigner-Ville time-frequency analysis method.

When smoothing is performed in both time and frequency directions, it is called smooth pseudo Wigner-Ville distribution (SPWVD), and its definition expression is as follows \cite{6923910}:
\begin{equation}
\begin{aligned}
W_x(t,f)&= \int g(u-t)\int h(\tau)x(t+\frac{\tau}{2})\\&~~~~~x^*(t-\frac{\tau}{2})e^{-j2\pi ft} d\tau du,\label{eq}
\end{aligned}
\end{equation}
where $g(u-t)$ and $h(\tau)$ are real symmetric window functions, $g(u-t)$ allows cross-terms oscillating parallel to the time axis perform smoothing (time smoothing), and $h(\tau)$ allows smoothing of cross-terms oscillating parallel to the frequency axis (frequency smoothing). Because smoothing is performed in both the time and frequency domains, the cross-interference terms of multi-component signals can be well suppressed. The WVD \cite{6530053}, PWVD \cite{Sun2018Bilinear}, and SPWVD time-frequency spectrograms of the frequency hopping signal are shown in Figure~\ref{fig4}. It can be clearly seen from the figure that SPWVD has a relatively good multi-component signal processing capability. 
\begin{figure}[htbp]
    \centering
    \includegraphics[width=0.48\textwidth]{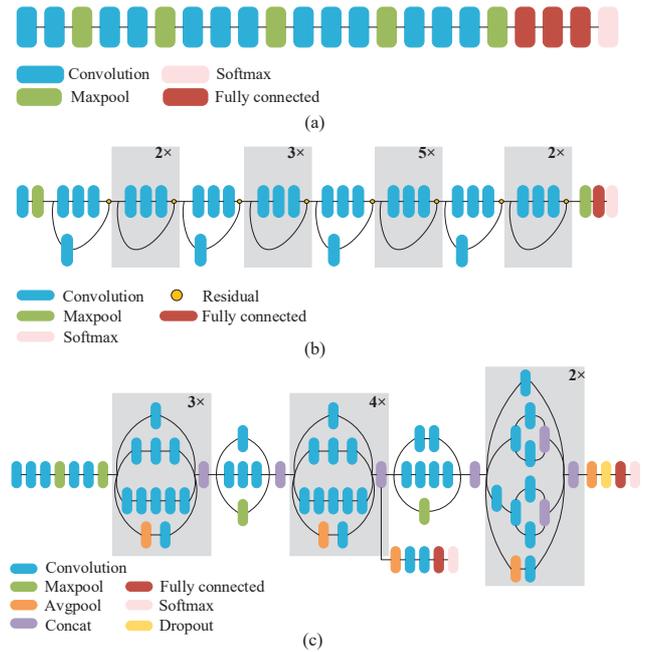}
    \caption{The network structure (compressed view). (a) is VGG16 network structure; (b) is ResNet50 network structure; (c) is InceptionV3 network structure.}
	\label{fig5}
\end{figure}

\section{Abnormal Signal Recognition with Convolutional Neural Networks}
One key ingredient of deep learning in image classification is the use of convolutional architectures \cite{8078730}. Convolutional neural networks (CNN) were proposed in the early days. As a representative of deep learning, CNN has great advantages in extracting discriminant and invariant features of inputs \cite{6472238}. The input of CNN is two-dimensional features (images), and the classification results are expressed in probabilistic form. Because CNN could extract features automatically and comprehensively, it is usually applied in the field of image recognition such as handwriting recognition and face detection \cite{1315150}.

\subsection{VGG16}
In 2014, the VGGNet architecture was proposed at the ImageNet Large Scale Visual Recognition Challenge (ILSVRC). The most important contribution is to increase the network depth (16 or 19 layers) by using a small convolution filter, which significantly improves image classification performance \cite{2013arXiv1312.6229S}. VGG16 has a strong generalization ability and is widely used as a relay feature extraction network for various detection network frameworks such as Fast-RCNN and SSD. Besides, the main advantage of using the VGG16 network structure is that it can be well extended to other data sets. Therefore, this paper extract features with VGG16 based on transfer learning.

The architecture of VGG16 is shown in Figure~\ref{fig5}. The convolutional layer receives an image with a size equal to 224×224×3 as input. Subsequently, the input image is propagated through a set of convolutional layers of 3×3, and the convolution step is taken as 1 pixel. The five largest pooling layers with a span equal to 2 are used for spatial pooling after several convolutional layers, and they are done on a 2×2 pixel window. After this set of conversion layers, there are three fully connected layers (FC) with channel sizes of 4096, 4096, and 1000. The softmax layer is the last layer, and its activation function is linear rectification function (ReLU): $f(x)=max(0,x)$.

\subsection{ResNet50}
Traditional convolutional networks may have problems such as information loss, and even cause gradients to disappear or explode, making very deep networks unable to train. ResNet (Residual Neural Network) solves this problem to a certain extent. The entire network only needs to learn the part of the difference between input and output, simplifying the learning objectives and difficulty. ResNet was proposed by four Chinese including Kaiming He and won the championship in the ILSVRC-2015 \cite{7780459}. Similar to VGG, 3×3 filters are used in this network. However, compared to the VGG network, ResNet has fewer filters and less complexity. Figure~\ref{fig5} illustrates a compressed view of ResNet50, which was used in this paper.

\subsection{InceptionV3}
Since 2014, building a deeper network model has become an important research direction of convolutional neural networks. The Google team developed a series of Inception models with the idea of deepening both the depth and width of the network model. Google Inception Net won first place in ILSVRC-2014. The network won with structural innovation. It uses a global average pooling layer to replace the fully connected layer, which greatly reduces the number of parameters. The network model is generally called Inception V1. In the subsequent Inception V2, the Batch Normalization method was introduced to speed up the convergence speed of training. In the Inception V3 model, the two-dimensional convolutional layer is split into two one-dimensional convolutional layers \cite{7780677}. It not only reduces the number of parameters but also reduces the over-fitting phenomenon. \begin{figure}[htbp]
    \centering
    \includegraphics[width=0.48\textwidth]{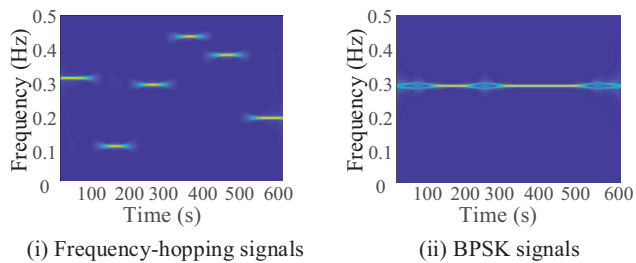}
    \caption{Time-frequency spectrograms of two normal signals.}
	\label{fig8}
\end{figure}
\begin{figure}[htbp]
    \centering
    \includegraphics[width=0.48\textwidth]{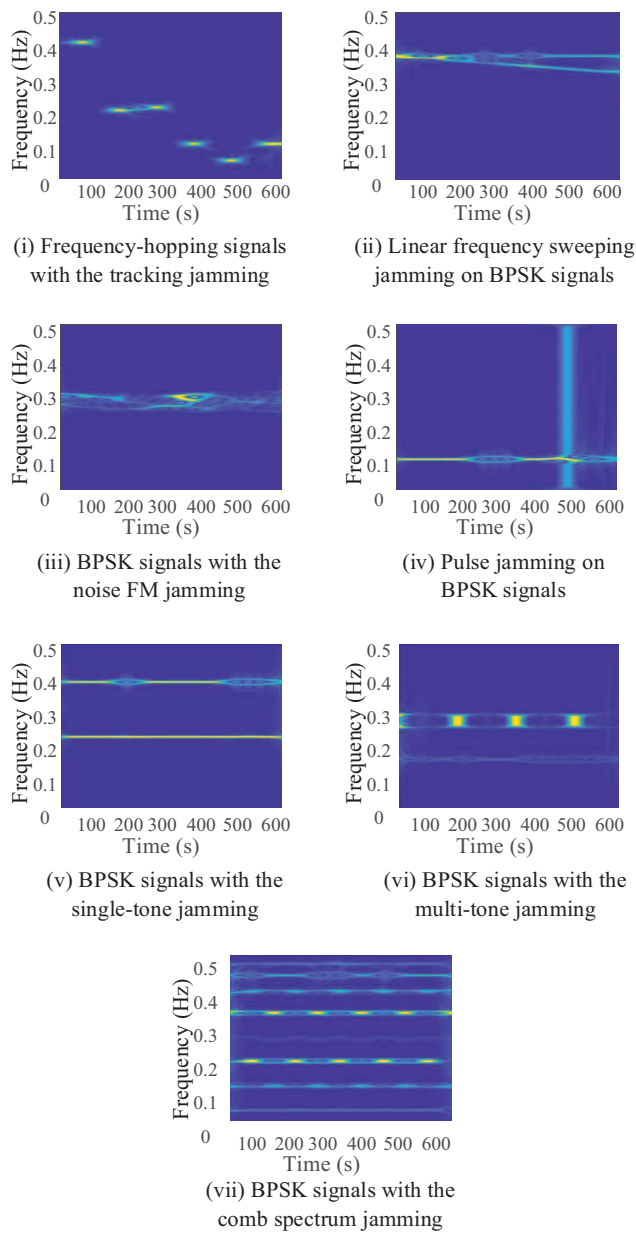}
    \caption{Time-frequency spectrograms of seven abnormal signals.}
	\label{fig9}
\end{figure}Therefore, compared to AlexNet, the network can learn deeper feature representations with fewer parameters, and it is much faster than VGG \cite{2016Xception}. Figure~\ref{fig5} illustrates a compressed view of InceptionV3 employed in this study.

To accommodate our task, we adjusted and modified a layer in VGG16, ResNet50, and InceptionV3 networks. The number of outputs of the softmax layer is changed to 9, which matches the number of recognition categories studied. To train better and faster, we divide the network into two parts (frozen and unfrozen) and perform cached training (passing the data through the frozen model once and then using them to train the unfrozen network) to avoid the difference between training and testing modes difference. Therefore, we have frozen the first 140 layers of the ResNet50 and InceptionV3 networks. The batch size is 32, the learning rate is 0.001, and 50 epochs have been run. The optimization algorithm is Adam. Besides, the method early stop and automatic attenuation of the learning rate are added to improve classification performance and prevent over-fitting.

\subsection{Implementation}
\begin{enumerate}
\item The communication signals are obtained through MATLAB simulation. The simulation parameters are shown in Table ~\ref{tab1}. Then, the MATLAB time-frequency analysis toolbox is used to perform a smooth pseudo Wigner-Ville time-frequency analysis on the signal to obtain a two-dimensional spectrogram of time-frequency distribution. The time-frequency spectrograms of normal signals are shown in Figure~\ref{fig8}, and the time-frequency spectrograms of abnormal signals are shown in Figure~\ref{fig9}.
\item Each image is labeled according to the style of signals.
\item The number of sampling for training samples and test samples is 600, and the jamming-to-signal ratio (JSR) is 5dB. Each type of signal generates training samples and test samples every 4dB between the signal-to-noise ratio (SNR) (-6dB, 10dB). The training samples have 2000 time-frequency spectrograms for each SNR, and each type of signal is 5×2000=10000, a total of 90,000; the test samples have 500 time-frequency spectrograms under each SNR, each type of signal 5×500=2500, a total of 22500. Besides, our simulation considers the influence of the Gaussian channel, Rayleigh fading channel, and frequency selective fading channel on the performance of the algorithm. In the Rayleigh fading channel, the standard deviation of the Rayleigh distribution is set to 0.5. For the frequency selective fading channels, we considered two paths. One is the original path without delay, and the other is the path with a delay of 50.
\item We also considered the influence of the JSR on the recognition performance of the algorithm in the Gaussian channel. Similarly, the number of sampling for training samples and test samples is 600, and the SNR is -2dB. Each type of signal generates training samples and test samples every 5dB between the JSR (-5dB, 10dB). The training samples have 2000 time-frequency spectrograms for each JSR, and each type of signal is 4×2000=8000, a total of 72,000; the test samples have 500 time-frequency spectrograms under each JSR, each type of signal 4×500=2000, a total of 18000.
\item The data set of the time-frequency spectrograms as the input is fed to deep learning networks for training using the Keras framework \cite{9010071}.
\item After a maximum of 563 training iterations, the trained model can be obtained. After training, the test samples are sent to the deep learning network to calculate the accuracy. The calculation result, i.e., the output of the network, is a probability vector indicating all possible categories, and the category with the highest probability is used as the result.
\end{enumerate}

\section{Simulation Results and Discussion}
\begin{figure}[htb]
    \centering
    \includegraphics[width=0.48\textwidth]{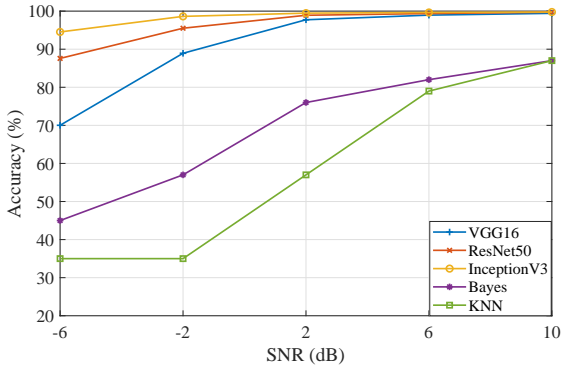}
    \caption{The recognition performance of different algorithms with JSR of 5dB the Gaussian channel.}
	\label{fig10}
\end{figure}

\begin{figure}[htb]
    \centering
    \includegraphics[width=0.48\textwidth]{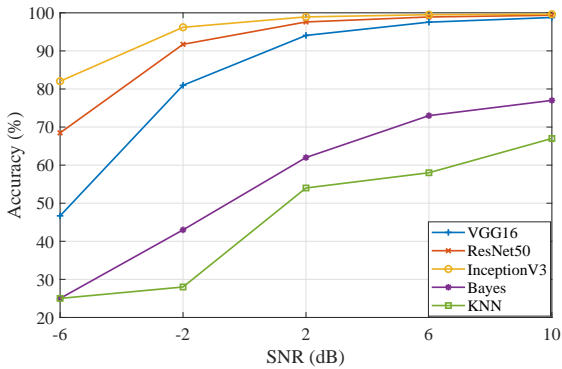}
    \caption{The recognition performance of different algorithms with JSR of 5dB in the Rayleigh fading channel.}
	\label{fig100}
\end{figure}

This section evaluates the performance of the proposed based on deep learning of the time-frequency spectrogram. Training is conducted with a computer equipped with an Nvidia GTX 1080 GPU. Figure~\ref{fig10} and Figure~\ref{fig100} evaluate and compare VGG16, ResNet50, and InceptionV3 with traditional machine learning image classification algorithms, which is the KNN algorithm and the Naive Bayes algorithm in the Gaussian channel and Rayleigh fading channel. 

\subsection{Impact of The Different Channels}
In the Gaussian channel as shown in Figure~\ref{fig10}, when the SNR is -6dB, the recognition rate of the deep learning algorithm is significantly better than that of the traditional machine learning algorithm. More specifically, even in the case of low SNR, the recognition rate of the InceptionV3 network reached more than 90\%. As the SNR increases, the recognition rate of various algorithms also increases. In the case of SNR is greater than 2dB, the accuracy of various deep learning algorithms gradually decreases. 
\begin{figure}[htb]
    \centering
    \includegraphics[width=0.48\textwidth]{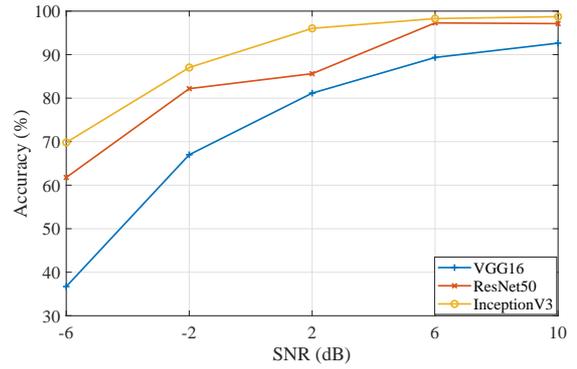}
    \caption{The recognition performance of different deep learning algorithms with JSR of 5dB in the frequency selective fading channel.}
	\label{fading}
\end{figure}
\begin{figure}[htb]
    \centering
    \includegraphics[width=0.48\textwidth]{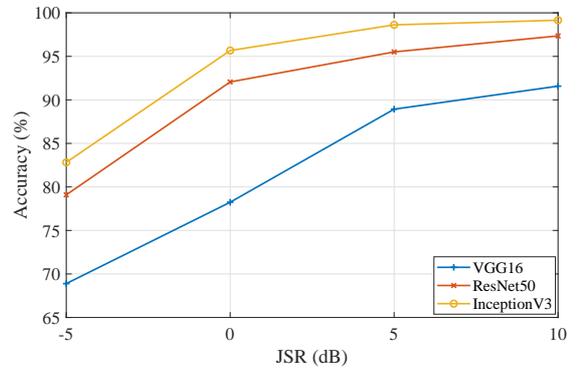}
    \caption{Performance comparison of different algorithms v.s. JSR.}
	\label{fig1000}
\end{figure}This is because when the SNR is high, the essential characteristics of the abnormal signal and the normal signal on the time-frequency spectrogram are becoming more and more obvious and intuitive. Therefore, it is getting easier for deep learning networks to automatically extract features, and it is easier to distinguish various types of signals.

In the Rayleigh fading channel as shown in Figure~\ref{fig100}, the recognition rates of deep learning algorithms and traditional machine learning algorithms both decrease. Especially the VGG16 network has the largest drop. However, the InceptionV3 network still performed the best, even in the case of low SNR, the recognition rate reached more than 80\%. This means that the InceptionV3 network has a certain degree of robustness and stability, and can adapt to worse situations.

In the frequency selective fading channel as shown in Figure~\ref{fading}, the performance of deep learning algorithms degrades more severely. The recognition rate of the VGG16 network drops to less than 40\% when the SNR is -6dB. ResNet50 and InceptionV3 networks performed well, and the recognition rate of InceptionV3 networks is 70\% and above. From the results, the InceptionV3 network can indeed adapt to a worse situation.
\begin{figure}[htbp]
    \centering
    \includegraphics[width=0.48\textwidth]{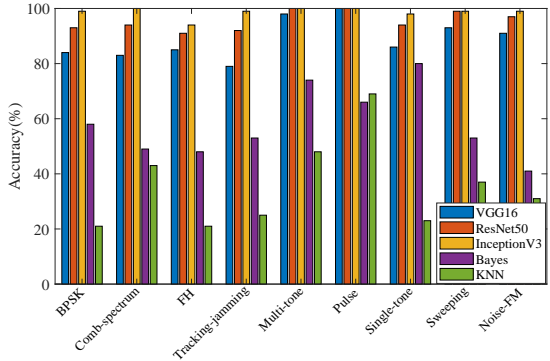}
    \caption{The recognition rate of different classification algorithms for each type of signal with SNR of -2dB in the Gaussian channel. (JSR is 5dB)}
	\label{fig11}
\end{figure}
\begin{figure}[htbp]
    \centering
    \includegraphics[width=0.48\textwidth]{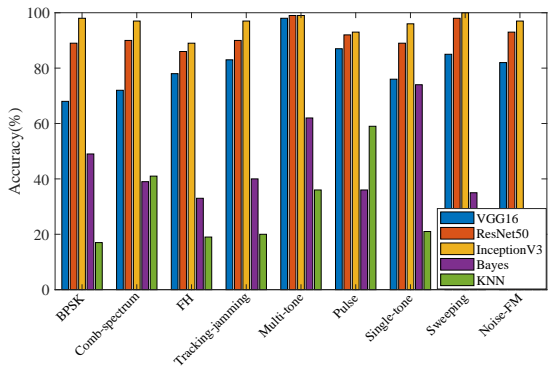}
    \caption{The recognition rate of different classification algorithms for each type of signal with SNR of -2dB in the Rayleigh fading channel. (JSR is 5dB)}
	\label{fig111}
\end{figure}
\begin{figure}[htbp]
    \centering
    \includegraphics[width=0.48\textwidth]{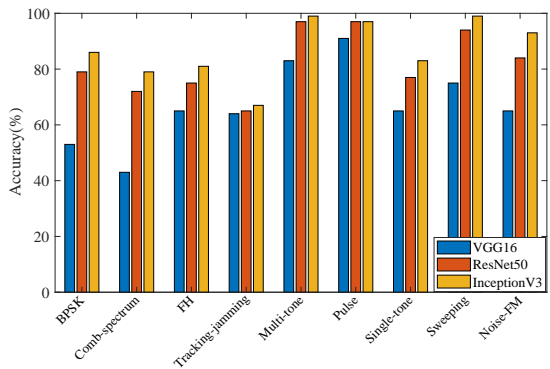}
    \caption{The recognition rate of different classification algorithms for each type of signal with SNR of -2dB in the frequency selective fading channel. (JSR is 5dB)}
	\label{fig1111}
\end{figure}
\subsection{Impact of The Different JSRs}
Figure~\ref{fig1000} shows the impact of JSR on the performance of different deep learning algorithms. When the JSR is -5dB, the identification of abnormal signals can be regarded as the identification of weak jamming signals, which is a challenging problem. From the results, we can observe that even in the case of low JSR, the performance is acceptable, and the recognition rate of the InceptionV3 network can reach more than 80\%. With the increase of JSR, the performance is also improved. Generally, when the jamming signal interferes with the normal communication signal, the power of the interference signal will be much greater than that of the normal communication signal. When the JSR is larger than 5dB, the recognition rate of various deep learning algorithms reaches 90\% and above. This shows that the abnormal signal recognition algorithm based on the time-frequency spectrogram deep learning can identify various abnormalities well.
\begin{figure}[htbp]
    \centering
    \includegraphics[width=0.45\textwidth]{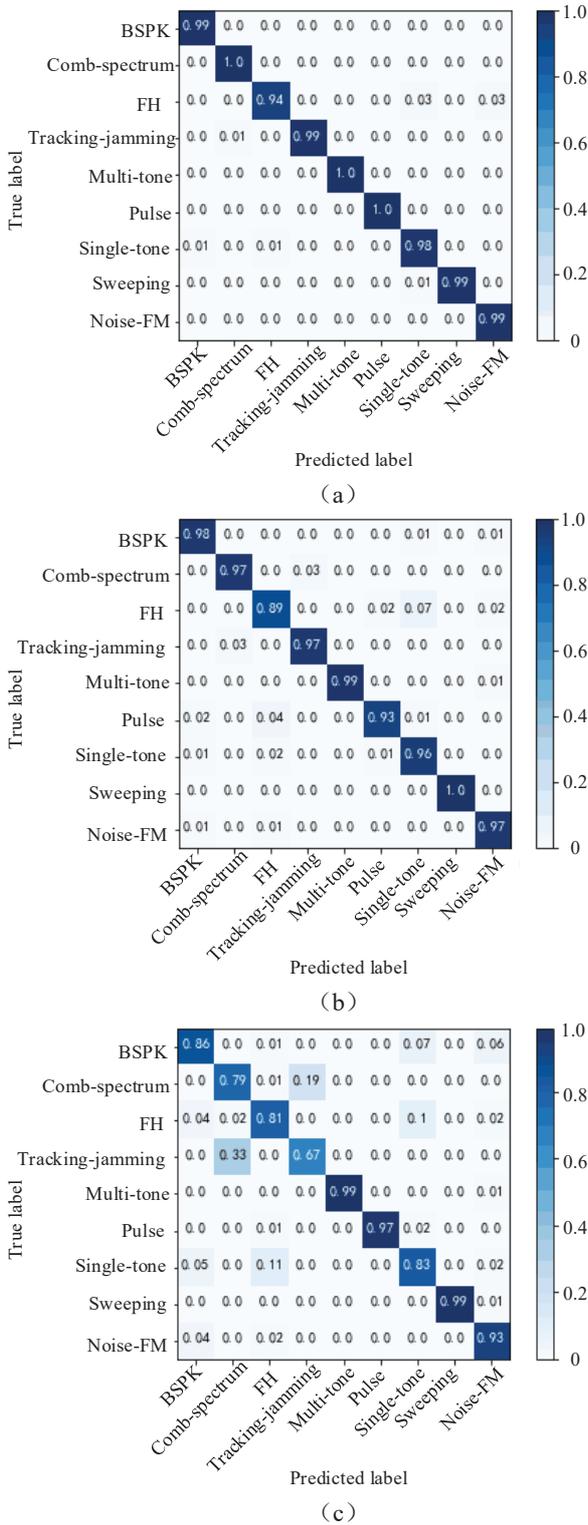}
    \caption{Confusion matrix of the recognition results, which obtained by using InceptionV3 at SNR of -2dB. (a) is confusion matrix of the Gaussian channel; (b) is confusion matrix of the Rayleigh fading channel. (c) is confusion matrix of the frequency selective fading channel.}
	\label{fig14}
\end{figure}

\subsection{Accuracy of Different Classification Methods for Each Signal}
Figure~\ref{fig11}, Figure~\ref{fig111}, and Figure~\ref{fig1111} show the recognition rate of different classification methods for each type of signal at SNR of -2dB in the three channels. From the results figures, the four signals of \emph{BPSK}, \emph{Comb-spectrum}, \emph{FH}, and \emph{Tracking-jamming} are more sensitive to channel changes, especially the \emph{Tracking-jamming} signal. The reason is that the \emph{Tracking-jamming} signal itself is very similar to the FH signal, and there is also the influence of noise and fading channels. For the ResNet50 and InceptionV3 network, the recognition accuracy of various types of abnormalities has reached or exceeded 80\% in the Gaussian channel and Rayleigh fading channel. Even in the frequency selective fading channel, the recognition rate of the InceptionV3 network for all types of signals except \emph{Tracking-jamming} signals still maintains 80\% and above. Compared with traditional machine learning algorithms, the proposed abnormal signal recognition algorithm based on time-frequency spectrogram deep learning has better performance. Among them, the accuracy of the fine-tuned InceptionV3 network is always the highest among these methods, especially in the low SNR regions.

Except for the recognition performance at different SNRs, the confusion matrix is also widely used in the multi-classification problem to analyze the recognition ability. As shown in Figure~\ref{fig14}, \emph{BPSK}, \emph{Comb-spectrum}, \emph{FH}, \emph{Tracking-jamming}, \emph{Multi-tone}, \emph{Pules}, \emph{Single-tone}, \emph{Sweeping} and \emph{Noise-FM} signals can be well recognized by the adjusted IncpetionV3 network. The model we proposed achieves excellent classification on the dataset. The high classification accuracy stresses the fact that extracting signal features from the time-frequency domain helps signals recognition.

\begin{table}[htbp]
\centering
\caption{Simulation parameters setting}
\renewcommand\tabcolsep{1.8pt}
\begin{tabular}{|c|c|c|ll}
\cline{1-3}
signal                           & \multicolumn{2}{c|}{parameters}&  &  \\ \cline{1-3}
\multirow{2}{*}{FH}              & $f_s$                    & 2000kHz               &  &  \\ \cline{2-3}
                                 & $f_1,f_2,f_3,f_4,f_5,f_6$& 100$\sim$2000kHz      &  &  \\ \cline{1-3}
\multirow{2}{*}{BPSK}            & symbol rate              & 100baud               &  &  \\ \cline{2-3}
                                 & carrier frequency        & 500$\sim$2500kHz      &  &  \\ \cline{1-3}
\multirow{2}{*}{\begin{tabular}[c]{@{}c@{}}Tracking-\\ jamming\end{tabular}}& $f_s$                    & 2000kHz               &  &  \\ \cline{2-3}
                                 & $f_1,f_2,f_3,f_4,f_5,f_6$& Same as FH signal     &  &  \\ \cline{1-3}
\multirow{2}{*}{Sweeping}        & $f_s$                    & 6000kHz               &  &  \\ \cline{2-3}
                                 & $f_0$                    & 0$\sim$2500kHz        &  &  \\ \cline{1-3}
\multirow{4}{*}{Noise-FM}        & $f_s$                    & 60000kHz              &  &  \\ \cline{2-3}
                                 & $K_{FM}$                   & 50000               &  &  \\ \cline{2-3}
                                 & center frequency         & 50000$\sim$250000kHz  &  &  \\ \cline{2-3}
                                 & background noise         & $\sigma_{n}^{2}$ =1            &  &  \\ \cline{1-3}
Pulse                            & time position            & random                &  &  \\ \cline{1-3}
\multirow{2}{*}{Single-tone}     & $f_s$                    & 8000kHz               &  &  \\ \cline{2-3}
                                 & carrier frequency        & 500$\sim$3500KHz      &  &  \\ \cline{1-3}
\multirow{2}{*}{Multi-tone}      & $f_s$                    & 8000kHz               &  &  \\ \cline{2-3}
                                 & $N$                      & 7                     &  &  \\ \cline{1-3}
\multirow{4}{*}{\begin{tabular}[c]{@{}c@{}}Comb-\\ spectrum\end{tabular}}   & $f_k$                    & 200$\sim$500kHz       &  &  \\ \cline{2-3}
                                 & $2\Delta f$              & 20$\sim$50kHz         &  &  \\ \cline{2-3}
                                 & $\alpha$                 & 0$\sim$0.5            &  &  \\ \cline{2-3}
                                 & $\beta$                  & 180$\sim$220          &  &  \\ \cline{1-3}
\end{tabular}
\label{tab1}
\end{table}
\begin{table}[htb]
\centering
\caption{Detect new types of abnormal signals (New type 0 denotes the power-law frequency modulation signal. New type 1 denotes the parabolic frequency modulation signal.)}
\renewcommand\tabcolsep{3.3pt}
\begin{tabular}{|c|c|c|c|c|c|}
\hline
SNR(dB)    & -6     & -2     & 2      & 6      & 10     \\ \hline
New type 0 & 53\%   & 89\%   & 79.4\% & 99.8\% & 100\%  \\ \hline
New type 1 & 88.8\% & 91.4\% & 87.8\% & 85.6\% & 87.2\% \\ \hline
\end{tabular}
\label{tab2}
\end{table}

\subsection{Detection of New types of Abnormal Signals}
Finally, we saved the trained InceptionV3 network model and tested it to detect new types of abnormal signals. One of the new types of abnormal signals is a power-law frequency modulation signal, where the degree of the power-law is randomly distributed between 0.15 and 0.5. The other is a parabolic frequency modulation signal, where the ordinates of the three points of the parabola are randomly distributed between 0.1 and 0.45. For each type of abnormal signal, 500 test samples are generated every 4dB between the signal-to-noise ratio (SNR) (-6dB, 10dB). Among them, JSR=5dB, and the channel is the Gaussian channel. Then, the test samples are fed to the trained model for testing. In the test, we used threshold filtering and the threshold is set to 0.95. When the confidence of each class of the inference result is less than this threshold, the input is considered to be a new abnormal signal. The simulation results are shown in Table ~\ref{tab2}. We can observe that our trained model has the capability to detect abnormalities when a new type of abnormal signal appears.

\section{Conclusion and Future Work}
In this paper, we have proposed an abnormal communication signal recognition algorithm based on time-frequency spectrograms and deep learning to deal with the non-stationary signal. The algorithm first visualizes the communication signal into a two-dimensional time-frequency spectrogram and converts the problem of abnormal communication signal recognition into an image classification problem. Then, make full use of the excellent capabilities of deep learning in the field of image classification to improve the intelligence level of signal recognition in complex electromagnetic environments. Simulation results show the effectiveness of this method even in the low SNR and low JSR environment. 

The important limitation lies in the fact that most of the existing classification methods are closed-set recognition. To classify effectively the unknown samples, the state-of-the-art open-set recognition is adopted  \cite{9040673}. In the future, more research efforts are deserved to identify the practical abnormal signals with open-set recognition combined with the proposed time-frequency analysis and deep learning methods.

\bibliographystyle{gbt7714-numerical}
\bibliography{ref}

\end{document}